\documentclass[12pt,preprint]{aastex}
\usepackage{epstopdf}
\shorttitle{Lead Oscillator Strengths}

\begin{document}
\title{Lifetimes and Oscillator Strengths for Ultraviolet Transitions in Singly-Ionized Lead}
\author{N.~Heidarian\altaffilmark{1}, R.~E.~Irving\altaffilmark{1}, A.~M.~Ritchey\altaffilmark{2}, S.~R.~Federman\altaffilmark{1}, D.~G.~Ellis\altaffilmark{1}, S.~Cheng\altaffilmark{1}, L.~J.~Curtis\altaffilmark{1}, and W.~A.~Furman\altaffilmark{3}}
\altaffiltext{1}{Department of Physics and Astronomy, University of Toledo, Toledo, OH 43606; negar.heidarianboroujeni@rockets.utoledo.edu, richard.irving@utoledo.edu, steven.federman@utoledo.edu, david.ellis@utoledo.edu, song.cheng@utoledo.edu, larry.curtis@utoledo.edu}
\altaffiltext{2}{Department of Astronomy, University of Washington, Seattle, WA 98195; aritchey@astro.washington.edu}
\altaffiltext{3}{Reed College, Portland, OR, 97202; afurman@reed.edu}

\begin{abstract}
We present the results of lifetime measurements made using beam-foil techniques on levels of astrophysical interest in Pb~{\sc ii} producing lines at 1203.6~\AA{} ($6s6p^{2}$\ $^{2}D_{3/2}$) and 1433.9~\AA{} ($6s^{2}6d$\ $^{2}D_{3/2}$). We also report the first detection of the Pb~{\sc ii}~$\lambda1203$ line in the interstellar medium (ISM) from an analysis of archival spectra acquired by the Space Telescope Imaging Spectrograph (STIS) onboard the \emph{Hubble Space Telescope} (\emph{HST}). The oscillator strengths derived from our experimental lifetimes for Pb~{\sc ii}~$\lambda\lambda1203$, $1433$ are generally consistent with recent theoretical results, including our own relativistic calculations. Our analysis of high-resolution \emph{HST}/STIS spectra helps to confirm the relative strengths of the Pb~{\sc ii}~$\lambda\lambda 1203$, $1433$ lines. However, the oscillator strength that we obtain for Pb~{\sc ii}~$\lambda1433$ ($0.321\pm0.034$) is significantly smaller than earlier theoretical values, which have been used to study the abundance of Pb in the ISM. Our revised oscillator strength for $\lambda1433$ yields an increase in the interstellar abundance of Pb of 0.43 dex over determinations based on the value given by Morton, indicating that the depletion of Pb onto interstellar dust grains is less severe than previously thought.
\end{abstract}


\section{INTRODUCTION}

Lead ($Z=82$) is the heaviest element thus far detected in the interstellar medium (ISM), and is one of only a handful of elements heavier than zinc ($Z>30$) that have been detected via ultraviolet interstellar absorption lines. Knowledge of the interstellar abundances of heavy elements yields insight into the processes of stellar nucleosynthesis responsible for the production of elements beyond the iron peak (i.e., the slow and rapid neutron-capture processes). Accurate gas-phase abundances for rare heavy elements can also provide us with a better understanding of the depletion processes by which atoms condense onto interstellar dust grains. With an ionization potential of 15.0~eV, Pb~{\sc ii} is the dominant form of lead in the neutral ISM. Observations of the $6s^{2}6p$~$^{2}P^{\rm o}_{1/2}$--$6s^{2}6d$~$^{2}D_{3/2}$ transition of Pb~{\sc ii} at 1433.906~\AA{} with the Goddard High-Resolution Spectrograph onboard the \emph{Hubble Space Telescope} (\emph{HST}) have been used to study the abundance of Pb in just two sight lines (Cardelli 1994; Welty et al.~1995). Additional detections of the Pb~{\sc ii}~$\lambda1433$ line are available in archival Space Telescope Imaging Spectrograph (STIS) data (A.~M.~Ritchey et al., in preparation). However, the total number of Pb~{\sc ii} detections in the ISM remains small, and, as a result, neither the abundance nor the depletion behavior of interstellar Pb are very well constrained.

In this paper, we report the first detection in the ISM of the $6s^{2}6p$~$^{2}P^{\rm o}_{1/2}$--$6s6p^2$~$^{2}D_{3/2}$ transition of Pb~{\sc ii} at 1203.616~\AA{}. We discovered this line by co-adding high-resolution \emph{HST}/STIS archival spectra for over 100 individual sight lines. A preliminary examination indicated that the Pb~{\sc ii}~$\lambda1203$ line was stronger than $\lambda1433$. However, an accurate oscillator strength ($f$-value) for the $\lambda1203$ line was not available. In the compilation by Morton (2000), for example, there is no $f$-value listed for the Pb~{\sc ii} transition at 1203.6~\AA{}. Theoretical transition probabilities for Pb~{\sc ii}~$\lambda1203$ and $\lambda1433$ were reported by Col\'{o}n \& Alonso-Medina (2001). However, these calculations, as well as those of Safronova et al.~(2005), yield an $f$-value for the $\lambda1433$ line that is significantly smaller than those from earlier theoretical efforts (Migdalek 1976; Cardelli et at.~1993), which were the basis for the $f$-value listed by Morton (2000) for Pb~{\sc ii}~$\lambda1433$. Here, we present the first experimentally determined oscillator strengths for the Pb~{\sc ii} transitions at 1203.6 and 1433.9~\AA{}, obtained from lifetime measurements made using beam-foil techniques.

The remainder of this paper is organized as follows. In Section 2, we describe the analysis of archival \emph{HST}/STIS spectra, which we use to derive empirically the relative $f$-values of the Pb~{\sc ii}~$\lambda1203$ and $\lambda1433$ lines. In Section 3, we discuss the lifetime measurements and the resulting oscillator strengths for several ultraviolet transitions in Pb~{\sc ii}. In Section 4, we present our own theoretical calculations of the Pb~{\sc ii} oscillator strengths, and compare these with our experimental results and with previous theoretical determinations. We summarize our conclusions in Section 5.

\section{ASTRONOMICAL OBSERVATIONS}

In an effort to search for previously undetected interstellar absorption lines in the far ultraviolet, we co-added high-resolution \emph{HST}/STIS archival spectra for large numbers of sight lines, producing high signal-to-noise ratio (S/N) composite spectra for different wavelength regions. Only spectra obtained with the E140H grating were considered. In addition, only sight lines that showed measurable interstellar absorption from the weak O~{\sc i} line at 1355.598~\AA{} were included. All spectra were aligned in velocity space, based on the position of the strongest O~{\sc i}~$\lambda1355$ absorption component, before co-adding, and were shifted to the laboratory rest frame. One absorption feature that was readily apparent in the composite spectra had a wavelength of 1203.616$\pm$0.003~\AA{}, precisely equal to that expected for the Pb~{\sc ii} line at 1203.616~\AA{}. Figure 1 shows a portion of the co-added spectrum, obtained from observations of 104 individual sight lines, showing the feature, which is detected with a significance of 9$\sigma$. We checked for other possible identifications for this feature using compilations for atomic lines (Morton 2000, 2003) and molecular lines (e.g., Abgrall et al.~1993a, 1993b; Morton \& Noreau 1994), but could not find any other reasonable identification other than Pb~{\sc ii}~$\lambda1203$.

Since the newly-discovered Pb~{\sc ii} line did not have an accurate $f$-value (prior to this study), we sought to determine the $f$-value empirically by comparing the $\lambda1203$ feature with Pb~{\sc ii}~$\lambda1433$. Of the 104 sight lines originally used to detect the $\lambda1203$ feature, only 34 also had high-resolution \emph{HST}/STIS spectra covering the Pb~{\sc ii}~$\lambda1433$ line. For these 34 sight lines, small spectral segments covering the 1203 and 1433~\AA{} regions were extracted from the STIS data, and were normalized by dividing each segment by its mean flux. The normalized spectra were then averaged, yielding a S/N of 100 at 1203~\AA{} and 240 at 1433~\AA{}. This procedure ensures that the column density associated with the resulting Pb~{\sc ii} absorption features corresponds to the average Pb~{\sc ii} column density for this particular set of sight lines, assuming the Pb~{\sc ii} lines are optically thin. While a more sophisticated weighting scheme could have been used to produce higher S/N composite spectra, the column densities associated with the Pb~{\sc ii}~$\lambda1203$ and $\lambda1433$ features resulting from such a scheme would no longer be a simple average, and would not necessarily be identical. Table 1 lists the background stars and the relevant information for the \emph{HST}/STIS data sets used to produce the average spectra.

The Pb~{\sc ii}~$\lambda1203$ and $\lambda1433$ features appearing in the average spectra were fit using the profile synthesis code ISMOD (see Sheffer et al.~2008). For these syntheses, the $f$-value of the $\lambda1433$ line was assumed to be $f_{1433}=0.321$, which is the value we obtain from our beam-foil experiments (see Section 3). The $f$-value of the $\lambda1203$ line was varied until the column densities resulting from the two fits were identical for the same Doppler $b$-value. With this procedure, we find $f_{1203}=0.752$ for $N$(Pb~{\sc ii})~=~$1.18\times10^{11}$~cm$^{-2}$ and $b=2.6$~km~s$^{-1}$. The fitted equivalent widths for the two lines are $W_{1203}=1.11\pm0.23$~m\AA{} and $W_{1433}=0.68\pm0.11$~m\AA{}, while the fitted wavelengths are 1203.616$\pm$0.003~\AA{} and 1433.903$\pm$0.003~\AA{}. The profile fits themselves are shown in Figure 2. From these results, we find that the ratio of $f$-values is $f_{1203}/f_{1433}=2.34\pm0.43$, regardless of the value assumed for $f_{1433}$. If the Pb~{\sc ii} lines are indeed optically thin, then the $f$-value ratio should be given by $f_{1203}/f_{1433}=(\lambda_{1433}/\lambda_{1203})^2W_{1203}/W_{1433}$. Use of this equation with the above equivalent widths yields an $f$-value ratio of 2.32, in agreement with the profile synthesis results.

\section{EXPERIMENTAL DETAILS}

\subsection{Measurements}

Lifetimes were measured using beam foil techniques at the Toledo Heavy Ion Accelerator (THIA; Haar et al.~1993; Schectman et al.~2000). Pb$^{+}$ ions were produced by a Danfysik Model 911A Universal Ion Source, extracted at 20 kV, magnetically analyzed, and then accelerated to a total kinetic energy of either 205 keV or 180 keV. The two different beam energies were used to study possible systematic effects (see, e.g., Federman et al.~1992). Using an electrostatic switchyard, magnetically selected ions were steered toward carbon foils with thicknesses ranging from 2.2 to 2.4 $\mu$g cm$^{-2}$. Typical beam currents were $\sim$100 nA to minimize foil breakage. Emission lines were analyzed with an Acton 1 m normal incidence vacuum ultraviolet monochromator. Post foil velocities were estimated to range from 0.391 to 0.392 mm ns$^{-1}$ for the 205 keV beam and from 0.419 to 0.420 mm ns$^{-1}$ for the 180 keV beam. Uncertainties in foil thickness and in the energy calibration of the accelerator will impact the precision of these post foil velocity determinations. We measured decay curves for the $6s^{2}6d$ $^{2}D_{3/2}$ level at 69,740 cm$^{-1}$ and the $6s6p^{2}$ $^{2}D_{3/2}$ level at 83,083 cm$^{-1}$ using the resonance lines at 1433.9 and 1203.6~\AA{}. Decay curves were also measured for the $6s6p^{2}$ $^{2}D_{5/2}$ level using the transition at 1335.20~\AA{}, and the $6s^{2}6d$ $^{2}D_{5/2}$ level was measured from the transition at 1822.05~\AA{}. All decay curves and spectra were normalized by means of an optical monitor.

Lifetimes were obtained through analysis of the decay curves. Two exponentials were required for the $6s^{2}6d$ $^{2}D_{3/2}$ level, since there is a repopulation coming from the cascade of the $5f$ $^{2}F_{5/2}$ level to the level of interest; we therefore employed the method of Arbitrarily Normalized Decay Curves (ANDC; Curtis et al.~1971) in the analysis, which enabled us to extract the lifetime of the primary decay for $6s^{2}6d$ $^{2}D_{3/2}$. Theoretical calculations indicate a lifetime for this cascade of 3.61 ns (Col\'{o}n \& Alonso-Medina 2001) and 5.55 ns (Alonso-Medina 1996) and the latter one is consistent with other experimental values (Gorshkov \& Verolainen 1985; Alonso-Medina 1997) as well as our ANDC analysis. For the $6s6p^{2}$ $ ^{2}D_{3/2}$ level, we did not find a noticeable cascade; the ANDC analysis gave the same primary lifetime as the one we obtained from a single-exponential fit. The lifetime for the $6s6p^{2}$ $^{2}D_{5/2}$ level was obtained by means of a single-exponential fit as well. Measuring the decay curve for this level (using the $\lambda1335.2$ line) was challenging since there are two possibly strong potential blends coming from resonance lines of C~{\sc ii} at 1334.5 and 1335.7~\AA{}. The C~{\sc ii}~$\lambda1335.7$ line is twice as strong as $\lambda1334.5$ (Morton 2003). Thus, to avoid the possible blending from C~{\sc ii}~$\lambda1335.7$, we took the measurements at 1334.9~\AA{}. The lifetime for the  $6s^{2}6d$ $^{2}D_{5/2}$ level was also obtained from a single-exponential fit using the $\lambda1822$ line. No evidence was found for a potential cascade in this line either. 

Finally, in order to derive oscillator strengths from lifetimes, we need the branching fractions for the transitions we have measured. We adopted the values from Col\'{o}n \& Alonso-Medina (2001) because the various branches span too large a range for our experimental techniques. For  $\lambda1433$, the branching fraction from Col\'{o}n \& Alonso-Medina (2001) is 0.797, while, for $\lambda1203$, it is essentially 1.000. For comparison, the values from our own theoretical calculations, described in Section~4, are 0.770 and 0.999 for $\lambda1433$ and $\lambda1203$, respectively. For $\lambda1335$ and $\lambda1822$, there are no significant additional decay channels. 

\subsection{Results}

Decay curves are shown in Figures 3, 4, 5 and 6 for $\lambda1433$, $\lambda1203$, $\lambda1335$ and $\lambda1822$ respectively. Reported lifetimes correspond to the weighted averages of the lifetimes obtained with the two beam energies. Uncertainties are derived from the weighted uncertainty of exponential fit errors and systematic errors (estimated from the range in lifetimes measured from the two different energies), added in quadrature.

The results for lifetime measurements are shown in Table 2 and the resulting oscillator strengths are given in Table 3. The values for $\lambda1796$ and $\lambda1449$ are derived from the measured lifetimes and the corresponding branching fractions from Col\'{o}n \& Alonso-Medina (2001). Previous theoretical lifetimes and oscillator strengths are also reported in Tables 2 and 3. Our experimental oscillator strengths for Pb~{\sc ii}~$\lambda\lambda 1203$, $1433$ yield an $f$-value ratio of $f_{1203}/f_{1433}=2.3\pm0.2$, which is in very good agreement with the ratio we obtain from our analysis of high-resolution \emph{HST}/STIS spectra (see Section 2).

As mentioned above, the decay of $\lambda1433$ is significantly affected by a cascade, whereas the $\lambda1203$ decay may be treated as essentially cascade-free. Our theoretical calculations, discussed in the next section, help to explain this; the $6s^{2}5f$ $^{2}F_{5/2}^{\rm o}$ level decays to $6s^{2}6d$ $^{2}D_{3/2}$ by means of a strong one-electron transition with a transition probability of $A\approx0.55$ ns$^{-1}$, whereas its decay to the $6s6p^{2}$ $^{2}D_{3/2}$ level is nominally a two-electron change, and thus relies on configuration mixing resulting in a smaller value of $A\approx0.025$ ns$^{-1}$. This helps to confirm our ability to detect the presence or absence of significant cascade effects in our measurements.

\section{THEORETICAL CALCULATIONS}

The ion Pb$^+$ has 81 electrons; its ground electronic configuration is [Xe]$4f^{14}$ $5d^{10}$ $6s^2$ $6p$, where [Xe] denotes the xenon-like core of 54 electrons (Earls \& Sawyer~1935). The ground state is designated as $6s^26p$ $^{2}P^{\rm o}_{1/2}$ in Russell-Saunders (LS) coupling notation. The lowest even-parity configurations are ${6s6p^2}$, $6s^{2}7s$ and ${6s^26d}$. The ${6s6p^2}$ configuration gives rise to four LS-coupled terms: $^{4}P$, $^{2}D$, $^{2}P$, and ${^{2}S}$. We have calculated the transition rates from the ground $6p$ to all these excited configurations, though we are especially interested in the resonance lines involving the lowest $^{2}D_{3/2}$ levels: {$6s^{2}6p$ $^{2}P^{\rm o}_{1/2}$ $\longrightarrow$ $6s^{2}6d$ $^{2}D_{3/2}$ \ ($\lambda1433$)} and {$6s^{2}6p$ $^{2}P^{\rm o}_{1/2}$ $\longrightarrow$ $6s6p^{2}$ $^{2}D_{3/2}$ \ ($\lambda1203$)}.

Several calculations have been made recently with varying results. The one most relevant for us here is that of Col\'{o}n \& Alonso-Medina (2001), who point out that the $^{2}D$ terms are subject to considerable interaction between the $6s6p^{2}$ and $6s^{2}6d$ configurations. Their calculation adopted the method of Cowan~(1981) in which relativistic Hartree-Fock (HFR) wavefunctions, configuration interactions in intermediate coupling, and a fit to observed energy levels are used to determine the mixing coefficients and transition probabilities. In another recent effort, Safronova et al.~(2005) presented a calculation of the $6s^{2}nd$ series, using relativistic many-body perturbation theory. Earlier theoretical efforts (Migdalek~1976; Cardelli et at.~1993), though consistent among themselves, produced $f$-values that are significantly larger than those from the two recent calculations for the $6s^{2}6d$ transitions.

We have done a multi-configuration Dirac Hartree Fock (MCDHF) calculation using GRASP2K (J{\"{o}}nsson et al.~2007; J{\"{o}}nsson et al.~2013), which is a general atomic structure program that includes both electron correlation and special relativity. Self-consistent multi-configuration equations using the Dirac-Coulomb Hamiltonian are solved to give the atomic state functions as linear combinations of relativistic configuration state functions (RCSFs) based on Dirac orbitals. Relativistic corrections, including the Breit interaction, are then added to the Hamiltonian and a relativistic configuration-interaction (RCI) calculation is performed in which the theoretical energy levels are produced as the eigenvalues of the resulting energy matrix.

We treat the Pb~{\sc ii} system as a core of 78 electrons with an additional 3 valence electrons in the $n=6$ shell. We begin with a reference solution using the three non-relativistic configurations ${6s^26p}$, ${6s6p^2}$, and $6s^{2}7s$, which contain the lowest energy levels. This involves 25 Dirac orbitals, all of which are varied to obtain a fully self-consistent solution. We then produce solutions of increasing complexity in two ways. (1) We add virtual orbitals of successively higher values of the $n$ and $l$ quantum numbers, with RCSFs formed by single and double electron substitutions from the reference solution. (2) We take account of core-valence correlation in a limited way by allowing single substitutions from the $5d$ orbital. In this way, we develop a basic solution with 32 Dirac orbitals, namely all those with $n+l\leq8$, and 5,590 RSCFs, most of which (5,173) involve the $5d^9$ group. This solution is optimized on the 15 lowest levels (11 even, 4 odd), and already shows significant interaction between the two $^{2}D_{3/2}$ levels.

Next, we add three layers of virtual orbitals, the outer layer being $6h$, $7g$, $8f$, $9d$, $10p$, $11s$. At each stage, we use single and double electron replacements, including the $5d^9$ subshell, and vary only the outer layer, keeping the other orbitals fixed. We now compute the even-parity levels separately from the odd. We finish with a pair of solutions having 61 relativistic orbitals, namely all those with $n + l \leq 11$. For the even-parity solution, we have 114,966 RCSFs, while, for the odd, we have 21,777. For the even solution, the radial functions for the outer orbitals are optimized on the lowest 9 even-parity levels, including the two $^{2}D$ terms, with standard (${2j+1}$) weighting. For the odd solution, we optimize on the 4 lowest levels, namely the $6p$ and $7p$ doublets. After obtaining these MCDHF solutions, we continue with the GRASP2K program suite, to do RCI calculations, including the Breit interaction, again one for each parity, repeating for each step as virtual orbitals are added. From these calculations, we get the lowest 13 energy eigenvalues. We then compute the oscillator strengths for the dipole transitions between these levels. Note that the relativistic treatment automatically includes spin-changing transitions (i.e., intercombination lines).

Our theoretical results are shown in Tables 2 and 3, along with our experimental results and previously published results.  The estimated precision for the theoretical $f$-values takes into account both the change in computed values as additional orbitals are added to the calculation, and the difference between the Babushkin (length) and the Coulomb (velocity) gauges.  The calculated oscillator strengths have been corrected for the differences between observed and calculated transition energies.  These energy differences are at most about $1\%$.  The calculated ratio of oscillator strengths is $f_{1203}/f_{1433}=4.20\pm0.23$.

\section{SUMMARY AND CONCLUSIONS}

We have discussed various determinations of the oscillator strengths for ultraviolet transitions in Pb~{\sc ii}. From an analysis of high-resolution \emph{HST}/STIS archival spectra, we find that the ratio of the $f$-values of the Pb~{\sc ii}~$\lambda1203$ and $\lambda1433$ lines is $f_{1203}/f_{1433}=2.34\pm0.43$. Our detection of the Pb~{\sc ii}~$\lambda1203$ feature in composite \emph{HST}/STIS spectra represents the first detection of this line in the ISM. Lifetimes were measured using beam-foil techniques for several levels in Pb~{\sc ii}. The resulting oscillator strengths for the Pb~{\sc ii}~$\lambda1203$ and $\lambda1433$ transitions are consistent with recent theoretical results, including our own relativistic calculations. Moreover, the $f$-value ratio from our beam-foil experiments ($f_{1203}/f_{1433}=2.3\pm0.2$) is in very good agreement with our result from astronomical spectra. However, the $f$-value we obtain for Pb~{\sc ii}~$\lambda1433$ ($0.321\pm0.034$) is significantly smaller than those from earlier theoretical calculations, which were the basis for the $f$-value listed by Morton (2000) for this transition. This is not surprising considering the complexities of this three-valence-electron system:  core-valence correlation, relativistic effects including the Breit interaction, and strong configuration mixing. Our GRASP2K calculation, which does address these complexities and is most consistent with our THIA results as shown in Table 3, is probably the best theoretical treatment of these transitions to date. The earlier calculations were based on one-active-electron approximations which did not adequately account for these complexities.

 Our revised $f$-value for Pb~{\sc ii}~$\lambda1433$ yields an increase in the gas-phase interstellar abundance of Pb of 0.43 dex over determinations based on the Morton (2000) value. With this revision, it appears that the depletion of Pb onto interstellar dust grains is not as severe as previously thought. However, additional detections of Pb~{\sc ii} in the ISM will ultimately be needed to fully understand the depletion behavior of this rare heavy element.

\acknowledgments
This work was supported by grant HST-AR-12123.001-A from the Space Telescope Science Institute. W.~A.~F.~participated in the Research Experiences for Undergraduates (REU) Program of the National Science Foundation under award number 1262810. A.~M.~R.~acknowledges support from the Kenilworth Fund of the New York Community Trust.

\bibliographystyle{plainnat}

\clearpage

\begin{deluxetable}{lccc|lccc}
\tablecolumns{8}
\tablewidth{0pt}
\tabletypesize{\scriptsize}
\tablecaption{\emph{HST}/STIS Data Sets}
\tablehead{\colhead{Star} & \colhead{Data Set} & \colhead{Exp.~Time} & \colhead{Cen.~Wave} & \colhead{Star} & \colhead{Data Set} & \colhead{Exp.~Time} & \colhead{Cen.~Wave} \\
\colhead{} & \colhead{} & \colhead{(s)} & \colhead{(\AA)} & \colhead{} & \colhead{} & \colhead{(s)} & \colhead{(\AA)} }
\startdata
HD 15137  & o5lh02 &  2957 & 1271 &     HD 121968 & o57r02 &  4457 & 1271 \\   
          & o6lz06 &  1200 & 1489 &               & o57r02 &  8383 & 1453 \\   
HD 23478  & o6lj01 &  2945 & 1271 &     HD 122879 & ob2611 &  1990 & 1234 \\   
          & o6lj01 &  1448 & 1453 &               & o5c037 &   360 & 1271 \\
HD 24190  & ob2604 &  1750 & 1234 &               & o5lh07 &  1416 & 1271 \\   
          & o6lj02 &  2940 & 1271 &               & obkr37 &   600 & 1271 \\
          & o6lj02 &  1448 & 1453 &               & o6lz57 &   600 & 1489 \\
HD 62542  & obik01 & 13298 & 1307 &     HD 124314 & o54307 &  1466 & 1271 \\   
          & obik02 & 13298 & 1343 &               & o6lz58 &   300 & 1489 \\
HD 90087  & obie15 &  2038 & 1271 &     HD 137595 & o6lj03 &   840 & 1271 \\   
          & o6lz32 &  1200 & 1489 &               & o6lj03 &   711 & 1453 \\   
HD 93205  & o4qx01 &  1200 & 1234 &     HD 147683 & o6lj06 &  2940 & 1271 \\   
          & o4qx01 &   780 & 1416 &               & o6lj06 &  1903 & 1453 \\   
HD 93222  & o4qx02 &  1680 & 1234 &     HD 148937 & obkr42 &   600 & 1271 \\
          & o4qx02 &  1140 & 1416 &               & o6f301 &   883 & 1380 \\   
HD 93843  & o5lh04 &  1396 & 1271 &     HD 177989 & o57r03 &  4557 & 1271 \\   
          & o6lz40 &   300 & 1489 &               & o57r03 &  8691 & 1453 \\   
HD 99857  & o54301 &  3921 & 1271 &               & o57r04 & 10356 & 1489 \\   
          & o6lz44 &  1200 & 1489 &     HD 195965 & o6bg01 &   415 & 1307 \\   
HD 99872  & ob2603 &  2130 & 1234 &               & o6bg01 &   300 & 1489 \\   
          & o6lj0i &  1890 & 1271 &     HD 201345 & ob2613 &  1746 & 1234 \\   
          & o6lj0i &  1260 & 1453 &               & o5c050 &   360 & 1271 \\
HD 99890  & obkj03 & 16857 & 1234 &               & obkr50 &   600 & 1271 \\
          & obkr3k &   600 & 1271 &               & o6359p &   360 & 1489 \\
          & obkj02 & 11990 & 1416 &     HD 202347 & o5g301 &   830 & 1271 \\   
HD 102065 & o4o001 &  7679 & 1234 &               & o5g301 &   620 & 1453 \\   
          & o4o001 &  1200 & 1416 &               & o5g301 &   900 & 1489 \\   
HD 103779 & o54302 &  1466 & 1271 &     HD 209339 & o5lh0b &  1416 & 1271 \\   
          & o63563 &   720 & 1489 &               & o6lz92 &  1200 & 1489 \\
HD 104705 & o57r01 &  2400 & 1271 &     HD 218915 & o57r05 &  2018 & 1271 \\   
          & o57r01 &  3320 & 1453 &               & o57r05 &  1300 & 1453 \\   
          & o57r01 &  2900 & 1489 &               & o57r05 &  1262 & 1489 \\   
HD 108639 & ob2601 &  3730 & 1234 &     HD 224151 & o54308 &  1496 & 1271 \\   
          & o6lj0a &  1860 & 1271 &               & o6lz96 &   300 & 1489 \\
          & o6lj0a &  1800 & 1453 &     HD 232522 & obkj06 & 17166 & 1234 \\
HD 110434 & o6lj0b &  2130 & 1271 &               & o5c08j &  1440 & 1271 \\
          & o6lj0b &  1020 & 1453 &               & obkj05 & 12776 & 1416 \\
HD 114886 & o6lj0d &  1560 & 1271 &     HD 303308 & o4qx04 &  2220 & 1271 \\   
          & o6lj0d &  1590 & 1453 &               & o4qx04 &  1560 & 1416 \\   
HD 116852 & o5c01c &   360 & 1271 & CPD$-$59 2603 & o4qx03 &  5160 & 1234 \\
          & o8na03 &  5722 & 1271 &               & o4qx03 &  3180 & 1416 \\
          & obkr1c &   600 & 1271 & \\
          & o63571 &   720 & 1489 & \\
\enddata
\end{deluxetable}

\clearpage

\begin{deluxetable}{lcccc} 
\tablecolumns{5}
\tablewidth{0pt}
\tabletypesize{\footnotesize}
\tablecaption{Lifetimes for the $6s^26d$ and $6s6p^{2}$ Levels}
\tablehead{\colhead{Level} & \multicolumn{4}{c}{$\tau$ (ns)} \\
\cline{2-5} \\
\colhead{} & \colhead{THIA\tablenotemark{a}} & \colhead{GRASP2K} & \colhead{Other Theory} & \colhead{Ref.} }
\startdata
$6d$ $^2D_{3/2}$ & 1.53(16)\tablenotemark{b} & 1.772(32) & 1.32, 1.16 & 1,2 \\
$6d$ $^2D_{5/2}$ & 4.13(32) & 6.084(358) & 13.75, 1.42 & 1,2 \\
$6s 6p^{2}$ $^{2}D_{3/2}$ & $0.58(2)$ & 0.386(20) & 0.215 & 1 \\
$6s 6p^{2}$ $^{2}D_{5/2}$ & $1.97(14)$ & 0.700(28) & 0.447 & 1 \\
\enddata
\tablenotetext{a}{Beam-foil results from the Toledo Heavy Ion Accelerator.}
\tablenotetext{b}{Numbers in parentheses are uncertainties, shown in the last digits.}
\tablerefs{(1) Col\'{o}n \& Alonso-Medina 2001 -- relativistic Hartree Fock (from quoted transition probability); (2) Safronova et al. 2005 -- all-order Dirac Fock.}
\end{deluxetable}

\begin{deluxetable}{lccccc} 
\tablecolumns{6}
\tablewidth{0pt}
\tabletypesize{\footnotesize}
\tablecaption{Oscillator Strengths for $^{2}P^{\rm o}$--$^{2}D$ Transitions}
\tablehead{\colhead{Transition} & \colhead{Wavelength} & \multicolumn{4}{c}{$f$-value} \\
\cline{3-6} \\
\colhead{} & \colhead{(\AA)} & \colhead{THIA\tablenotemark{a}} &  \colhead{GRASP2K} & \colhead{Other Theory} & \colhead{Ref.} }
\startdata
$6p$ $^2P^{\rm o}_{1/2}$--$6d$ $^2D_{3/2}$ & 1433.9 & 0.321(34)\tablenotemark{b} & 0.268(5) & 0.869, 0.86, 0.372, 0.4518 & 1,2,3,4 \\
$6p$ $^2P^{\rm o}_{3/2}$--$6d$ $^2D_{3/2}$ & 1796.67 & 0.064(7) & 0.063(1) & 0.100, 0.074, 0.06255 & 1,3,4 \\
$6p$ $^2P^{\rm o}_{3/2}$--$6d$ $^2D_{5/2}$ & 1822.05 &  0.179(14)& 0.123(7) & 0.880, 0.054, 0.5024 & 1,3,4 \\
\\
$6p$ $^{2}P_{1/2}$--$6s 6p^{2}$ $^{2}D_{3/2}$ & 1203.62 & 0.75(3) & 1.124(58) & 2.02 & 3 \\
$6p$ $^{2}P_{3/2}$--$6s 6p^{2}$ $^{2}D_{3/2}$ & 1449.35 & 0.0001(0) & 0.0011(3) & 0.0003 & 3 \\
$6p$ $^{2}P_{3/2}$--$6s 6p^{2}$ $^{2}D_{5/2}$ & 1335.20 & 0.204(14) & 0.573(23) & 0.897 & 3 \\
\enddata
\tablenotetext{a}{Beam-foil results from the Toledo Heavy Ion Accelerator.}
\tablenotetext{b}{Numbers in parentheses are uncertainties, shown in the last digits.}
\tablerefs{(1) Migda{\l}ek 1976 -- relativistic; (2) Cardelli et al. 1993 -- Coulomb approximation with core polarization; (3) Col\'{o}n \& Alonso-Medina 2001 -- relativistic Hartree Fock (from quoted transition probability); (4) Safronova et al. 2005 -- all-order Dirac Fock.} 
\end{deluxetable}

\clearpage

\begin{figure}
\centering
\includegraphics[width=0.9\textwidth]{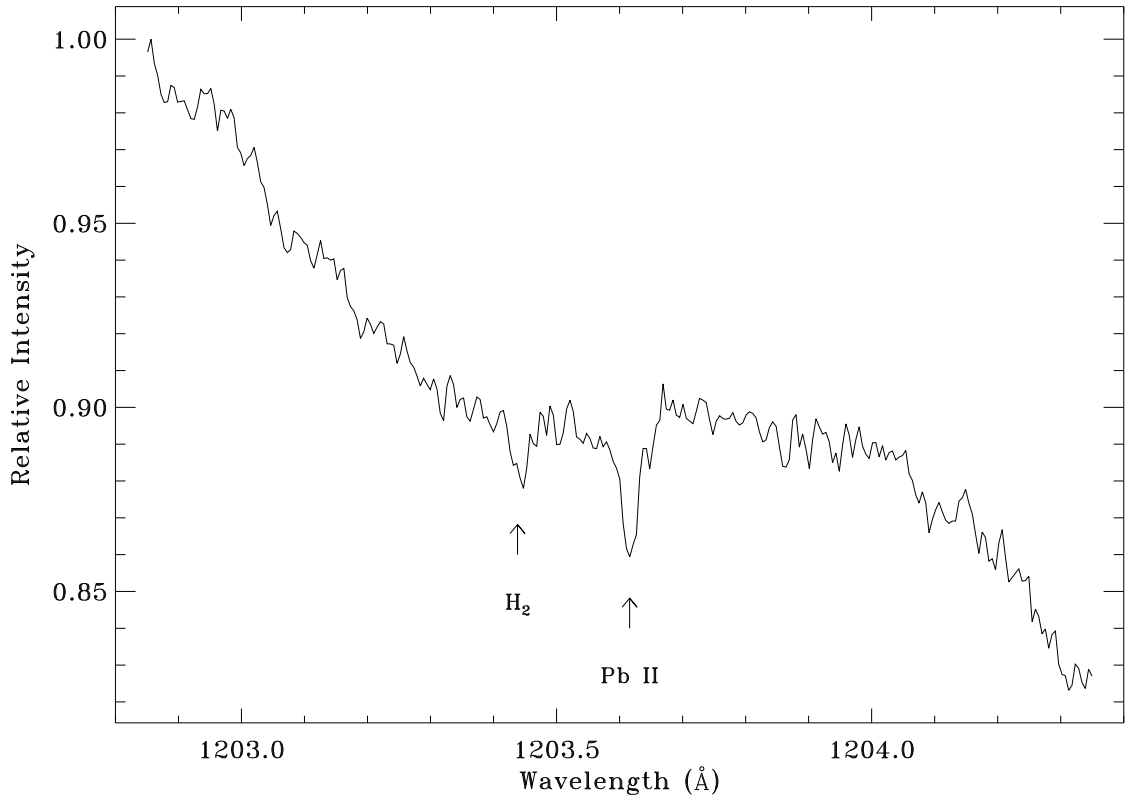}
\caption[]{Composite high-resolution \emph{HST}/STIS spectrum, obtained by co-adding data for 104 individual sight lines, covering the region surrounding the Pb~{\sc ii} line at 1203.616~\AA{}. The Pb~{\sc ii} feature is detected at the 9$\sigma$ level in this co-added spectrum. The other detected feature near 1203.44~\AA{} is a line of vibrationally-excited H$_2$, which is strong in some sight lines.}
\end{figure}

\begin{figure}
\centering
\includegraphics[width=0.7\textwidth]{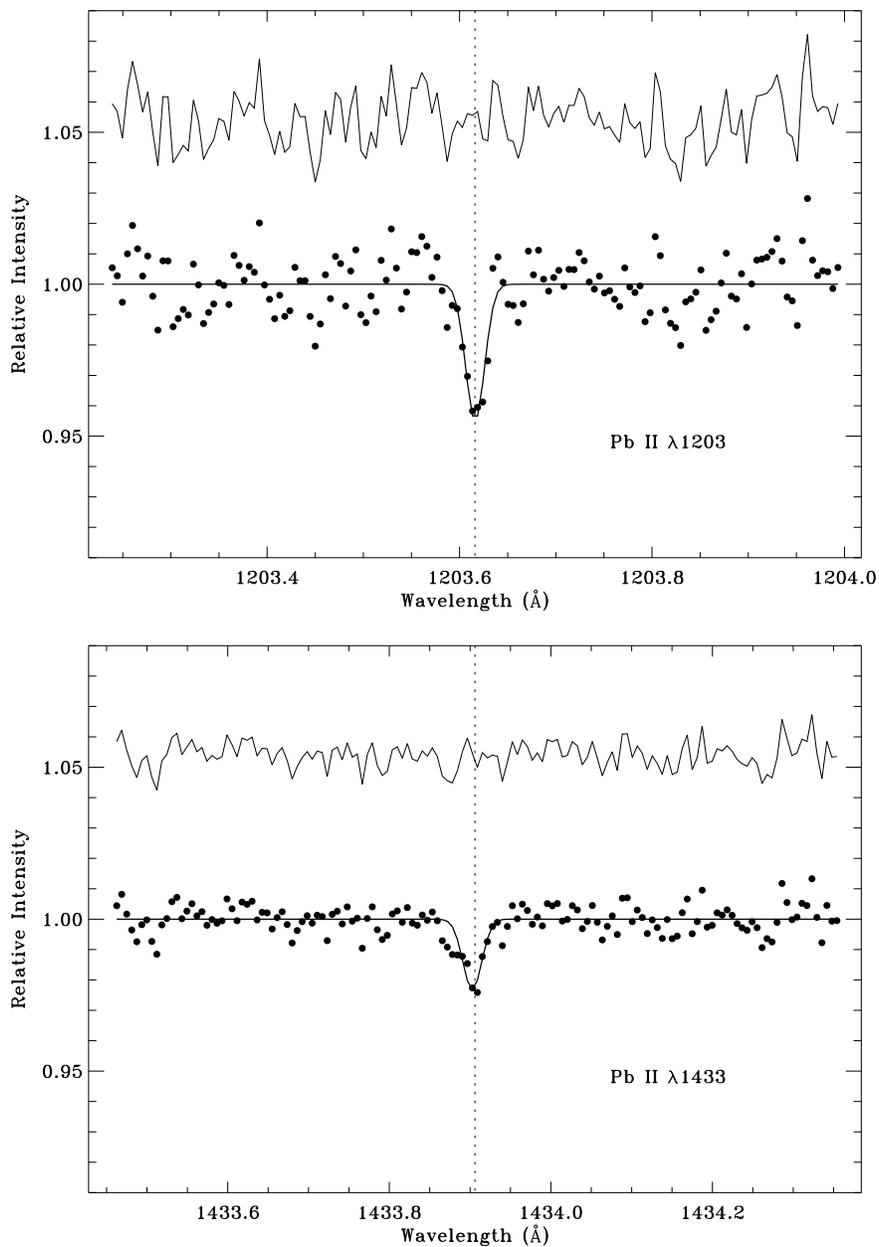}
\caption[]{Profile synthesis fits to the Pb~{\sc ii}~$\lambda1203$ and $\lambda1433$ features appearing in the average spectra for 34 sight lines with high-resolution \emph{HST}/STIS data covering both wavelength regions. Synthetic profiles are shown as solid lines passing through data points that represent the observed spectra. Residuals are plotted above each fit. The vertical dotted lines mark the expected positions of the Pb~{\sc ii} absorption features.}
\end{figure}

\begin{figure}[ht]
\centering
\includegraphics[width=0.7\textwidth]{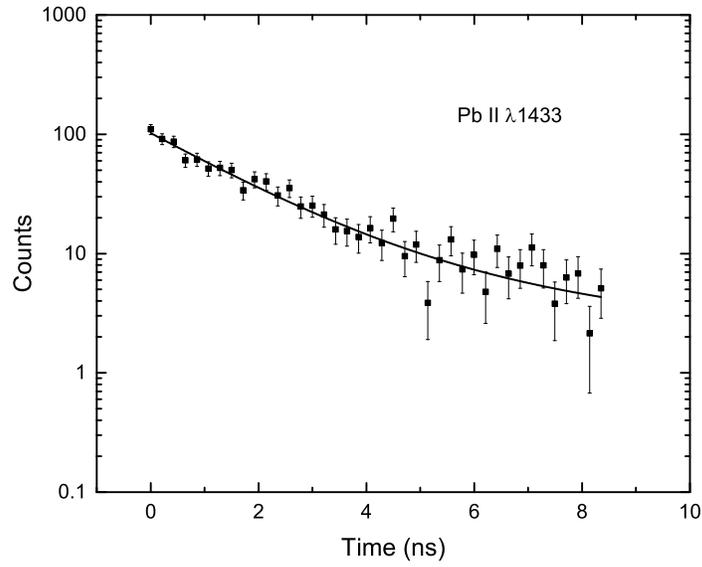}
\caption{Decay curve for the $\lambda1433$ line for a beam energy of 205 keV. The post-foil velocity ranges from $0.391$ mm ns$^{-1}$ to $0.392$ mm ns$^{-1}$ at this energy; these values were used to convert the foil position to the time.}
\end{figure}

\begin{figure}[ht]
\centering
\includegraphics[width=0.7\textwidth]{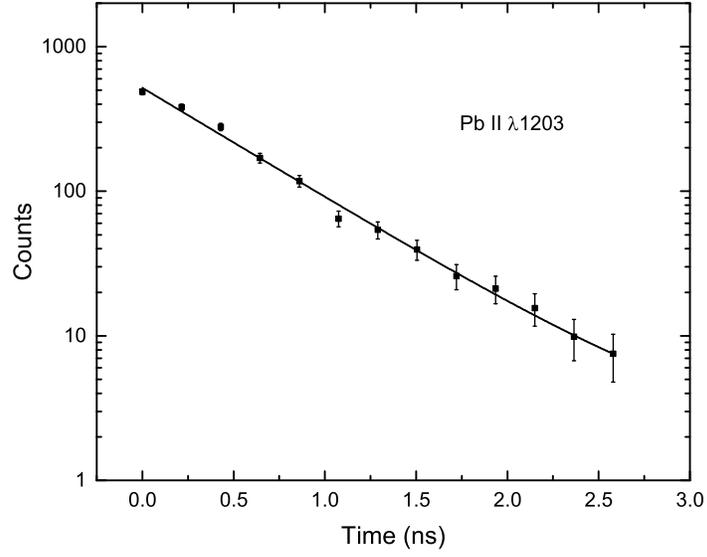}
\caption{Same as Fig. 3 for the $\lambda1203$ line for a beam energy of $205$ keV with a single-exponential fit as shown by the solid curve.}
\end{figure}

\begin{figure}[ht]
\centering
\includegraphics[width=0.7\textwidth]{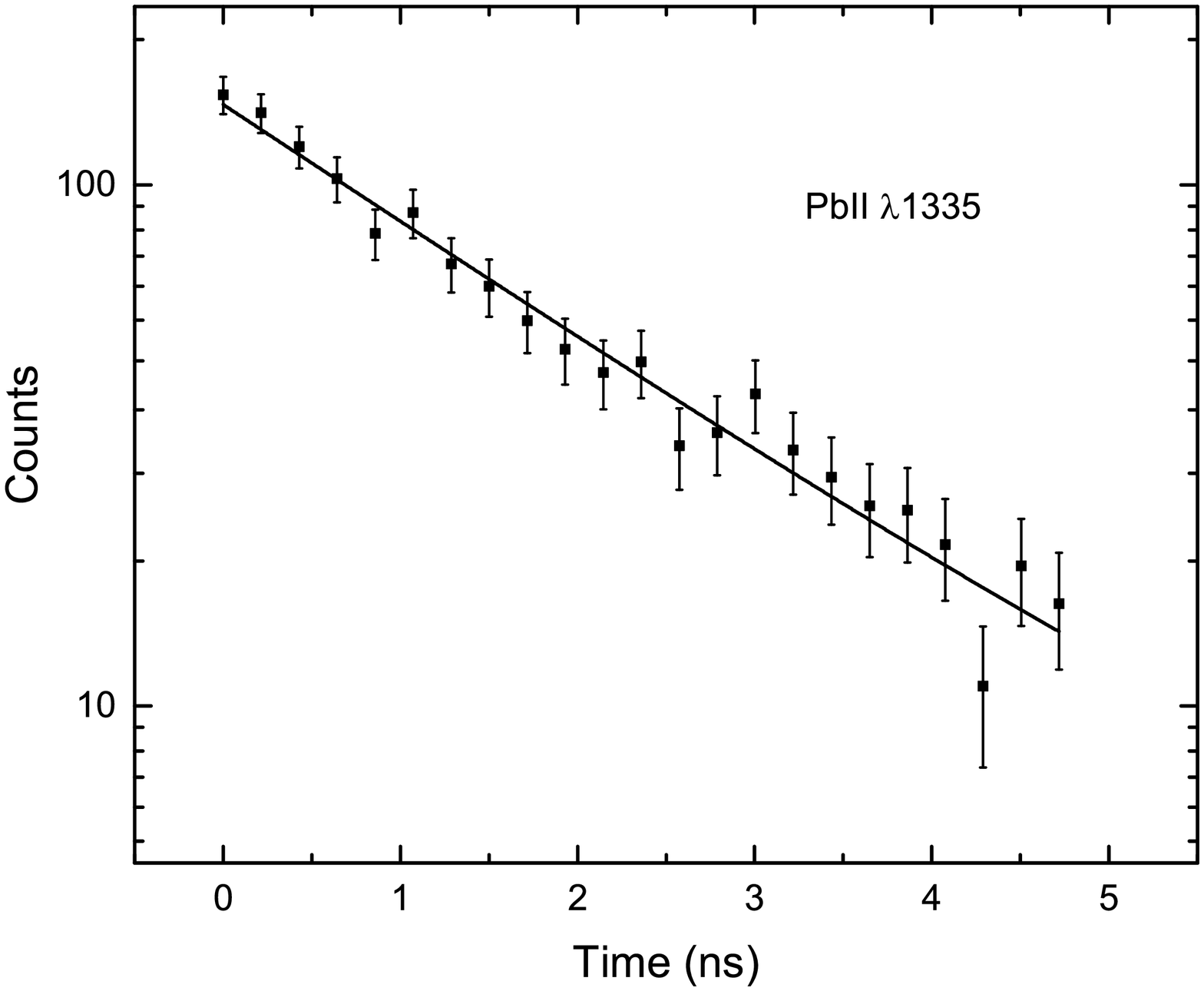}
\caption{Same as Fig. 3 except for the $\lambda1335$ line.}
\end{figure}

\begin{figure}[ht]
\centering
\includegraphics[width=0.7\textwidth]{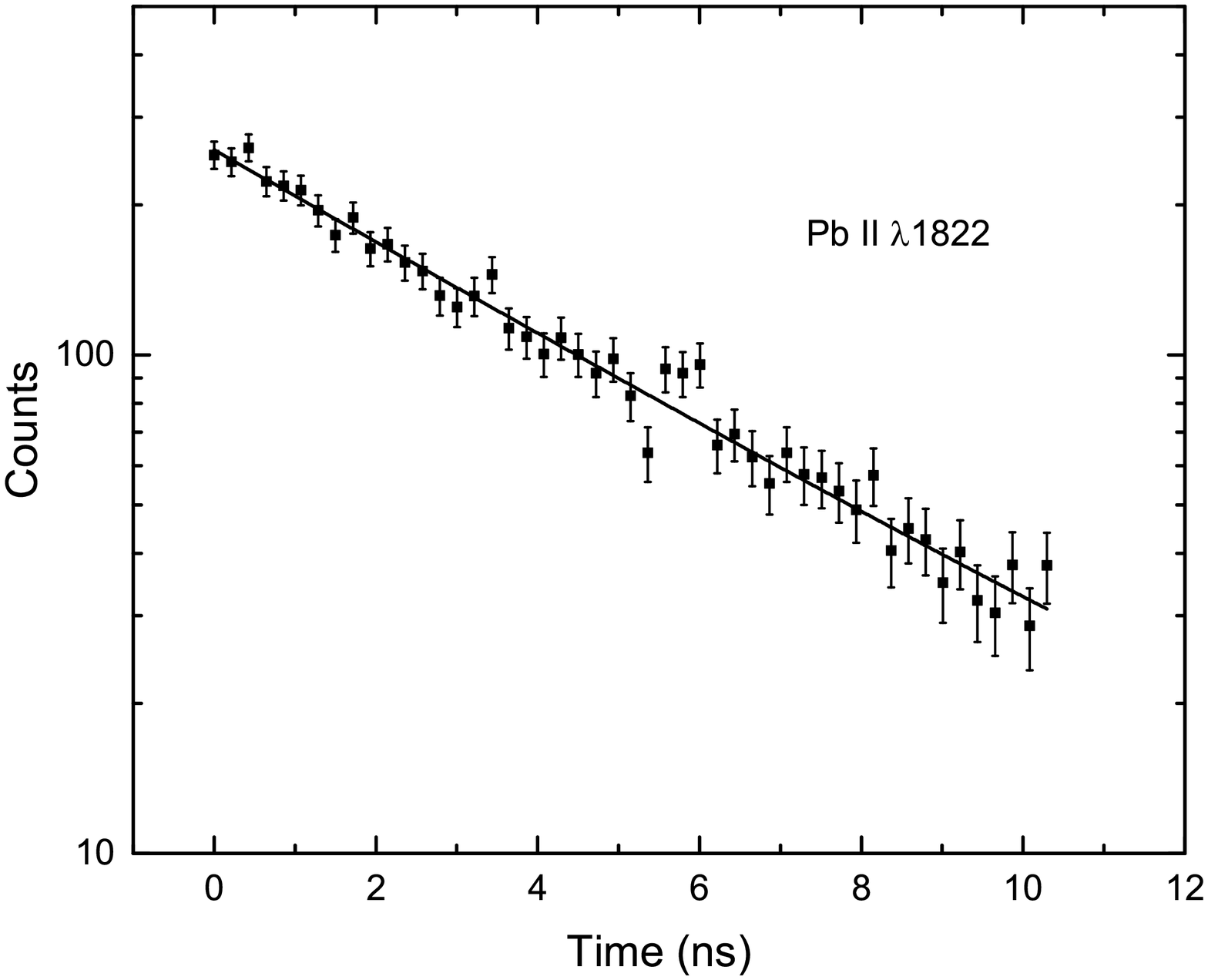}
\caption{Same as Fig. 4 except for the $\lambda1822$ line. }
\end{figure}

\end{document}